\documentclass{iaus}
\usepackage{graphicx}

\title[The image jets modeling of gravitationally lensed sources] {The image jets modeling of gravitationally lensed sources}

\author[T.Larchenkova, A.Lutovinov, N.Lyskova] 
{Tatiana Larchenkova$^1$,
Alexander Lutovinov$^2$,
Natalya Lyskova$^3$}

\affiliation{$^1$ASC of P.N.Lebedev Physical Institute, Profsoyuznaya str. 84/32,
Moscow 117997, Russia \\ email: {\tt tanya@lukash.asc.rssi.ru} \\
$^2$Space Research Institute, Profsoyuznaya str. 84/32, Moscow 117997, Russia \\
$^3$Moscow Physical-Technical Institute, Institutskyi per., 9, Dolgoprudnyi 141700, Russia \\[\affilskip]}

\pubyear{2010}
\volume{275} 
\jname{Jets at all Scales}
\editors{G.E.Romero, R.A.Sunyaev \& T.Belloni, eds.}
\begin{document}

\maketitle

\begin{abstract} 

The jets image modelling of gravitationally lensed sources have been
performed.  Several basic models of the lens mass distribution were
considered, in particular, a singular isothermal ellipsoid, an isothermal
ellipsoid with the core, different multi-components models with the galactic
disk, halo and bulge.  The obtained jet images were compared as with each
other as with results of observations.  A significant dependence of the
Hubble constant on the model parameters was revealed for B0218+357, when the
circular structure was took into account.

\end{abstract}

It is known, that due to the gravitational lensing of a compact object with
a relativistic jet the multiple images of the object itself and of the jet
can arise. Such gravitationally lensing systems are observed and the most
bright from them are PKS1830-211 (Nair et al., 1993) and B0218+357 (Patnaik
et al., 1993). The modelling of images of such sources and study their
temporal behavior are began actual due to forthcoming launches of cosmic
radiointerferometers.

\begin{figure}[b]
\begin{center}
\hbox{
 \includegraphics[width=4.2cm,bb=17 51 577 610]{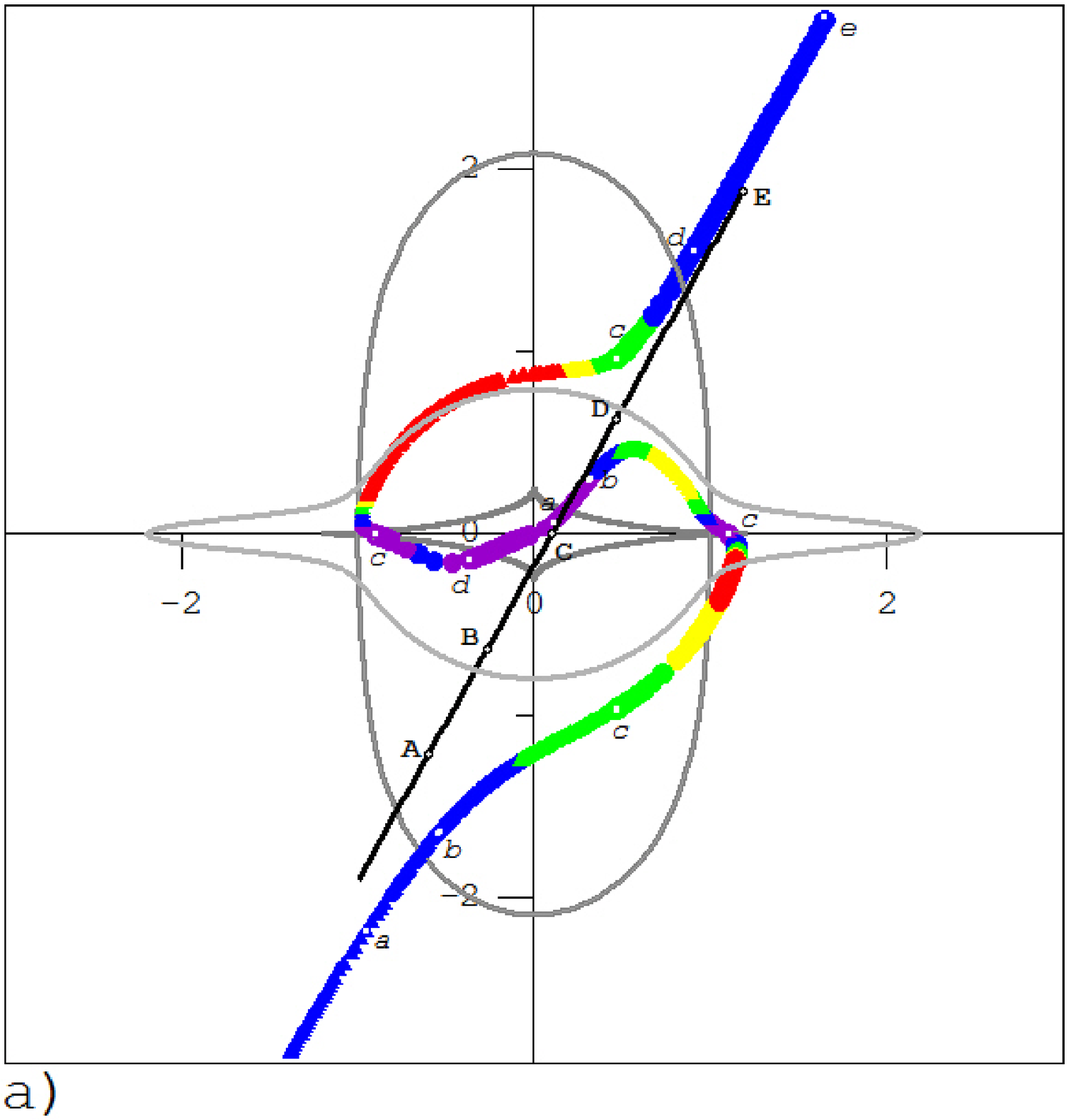}
 \includegraphics[width=4.2cm,bb=17 51 577 610]{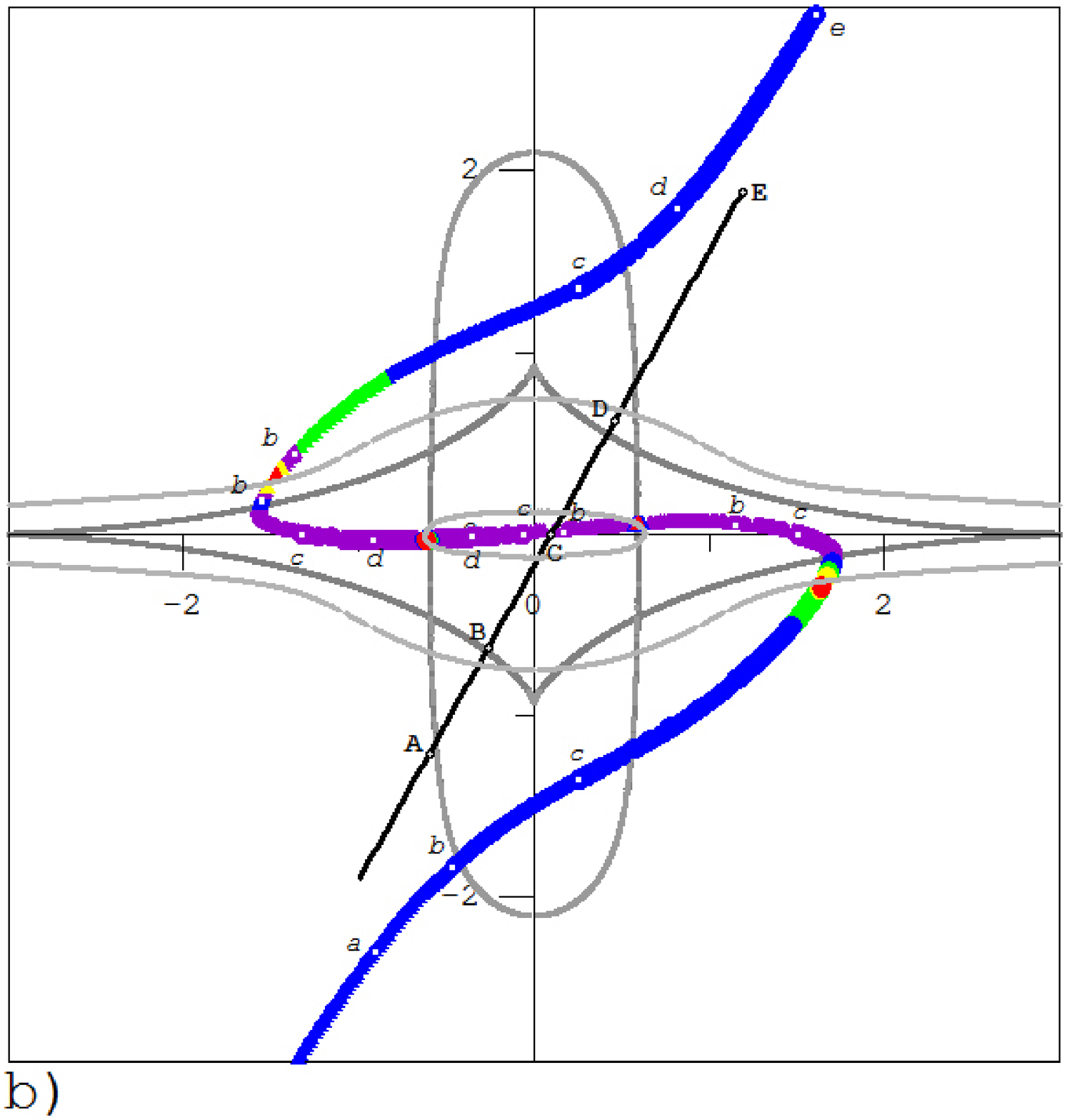}
 \includegraphics[width=4.2cm,bb=17 51 577 610]{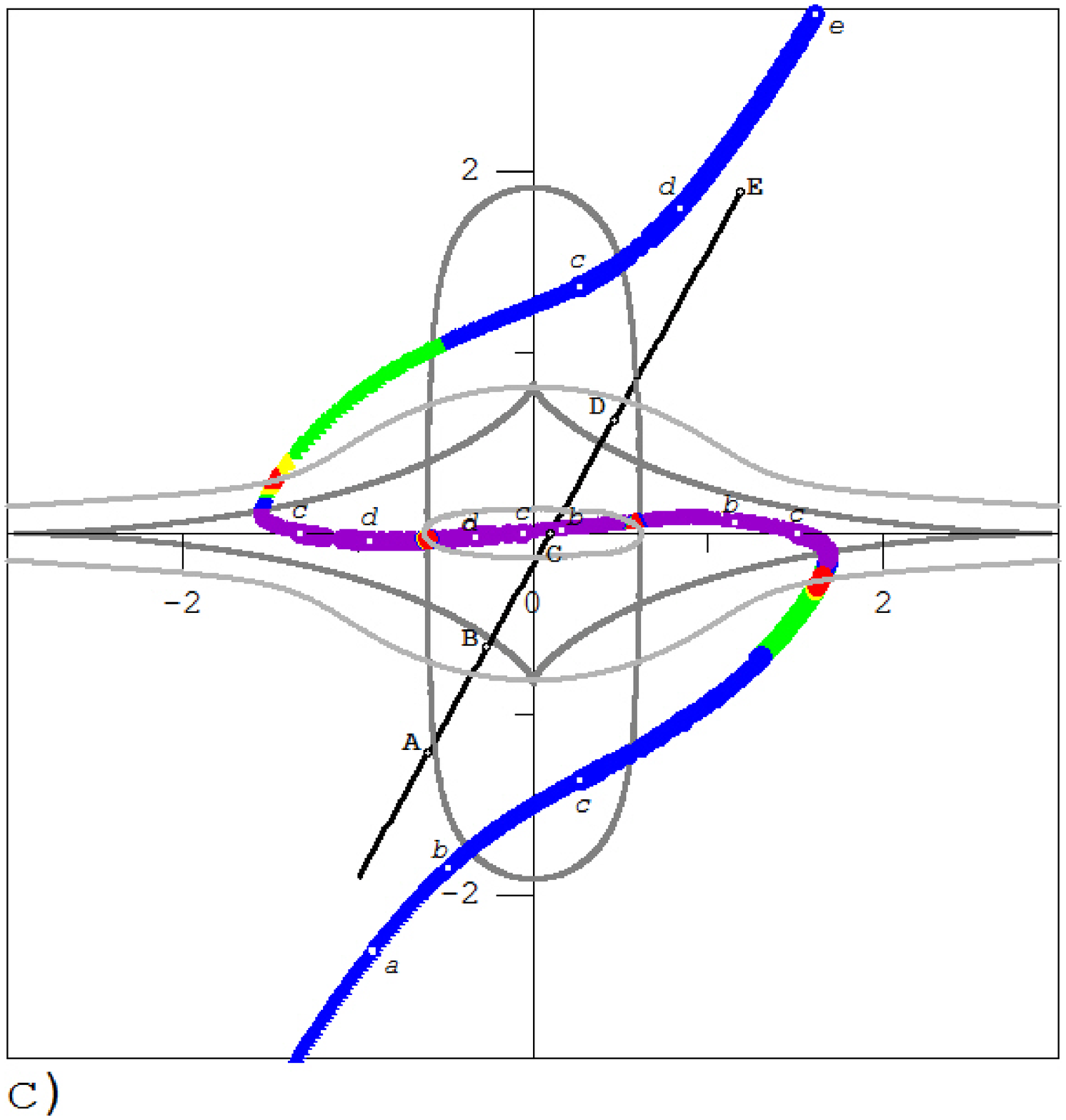}
 }

\end{center} \caption{Mapping of a gravitationally lensed jet for three
models (I - a, II - b, III - c).  Designations are: gray - caustics,
light-gray - critical curves, black line - jet.  Jets images depending on
the magnification $M$ are shown by magenta ($0<|M|<1$), blue ($1<|M|<3$),
green ($3<|M|<7$), yellow ($7<|M|<10$) and red ($|M|>10$).  The lower case
letters on the jet images correspond to the same capital letters on the jet
itself (for details see Larchenkova et al. 2011).}\label{fig1}

\end{figure}

For the case of the lensing on the spiral galaxy we considered several
multicomponents models: Model I -- disk and softened halo, approximated by
the isothermal ellipsoid located in the singular isothermal halo of a dark
matter; Model II - disk, approximated by the Kuzmin disk (Kuzmin, 1956) in
the isothermal halo; Model III - disk and bulge, approximated by the Kuzmin
disk, in the isothermal halo (Keeton, Kochanek 1998).

Depending on the disk axes ratio $q_{3d}$, its characteristics size $a_d$
and jet parameters, the different geometry of caustics and jet images was
obtained. Some typical results are presented in Fig.\ref{fig1} for different
lens models. It is clearly seen that the inclusion of the low-mass bulge
(similar to the Milky Way one) does not change significantly the picture of
the jet lensing (see Fig.1 b, c).

There is a general question: is it possible to obtain a circular-like
structure by the jet lensing?  The one of answers is that: to obtain a
circular-like structure it is necessary to cross the tangential caustics by
the jet near tangentially to their cusps. An example of the appearance of
such a structure are shown in Fig.2(left).

\begin{figure}[]

\hspace{15mm}\hbox{
 \includegraphics[width=5.5cm,bb=15 15 577 577]{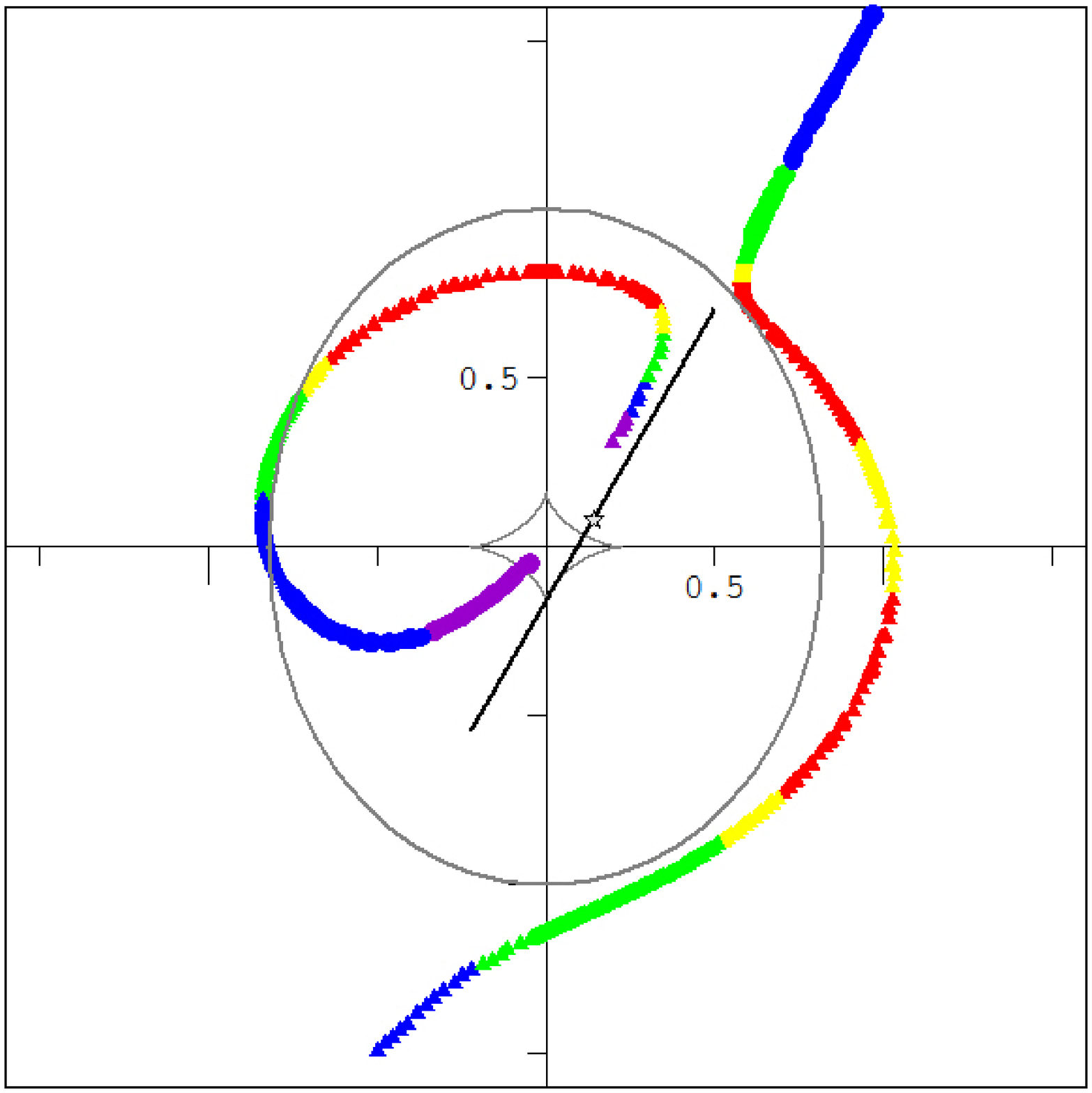}
 \includegraphics[width=5.5cm,bb=15 15 577 577]{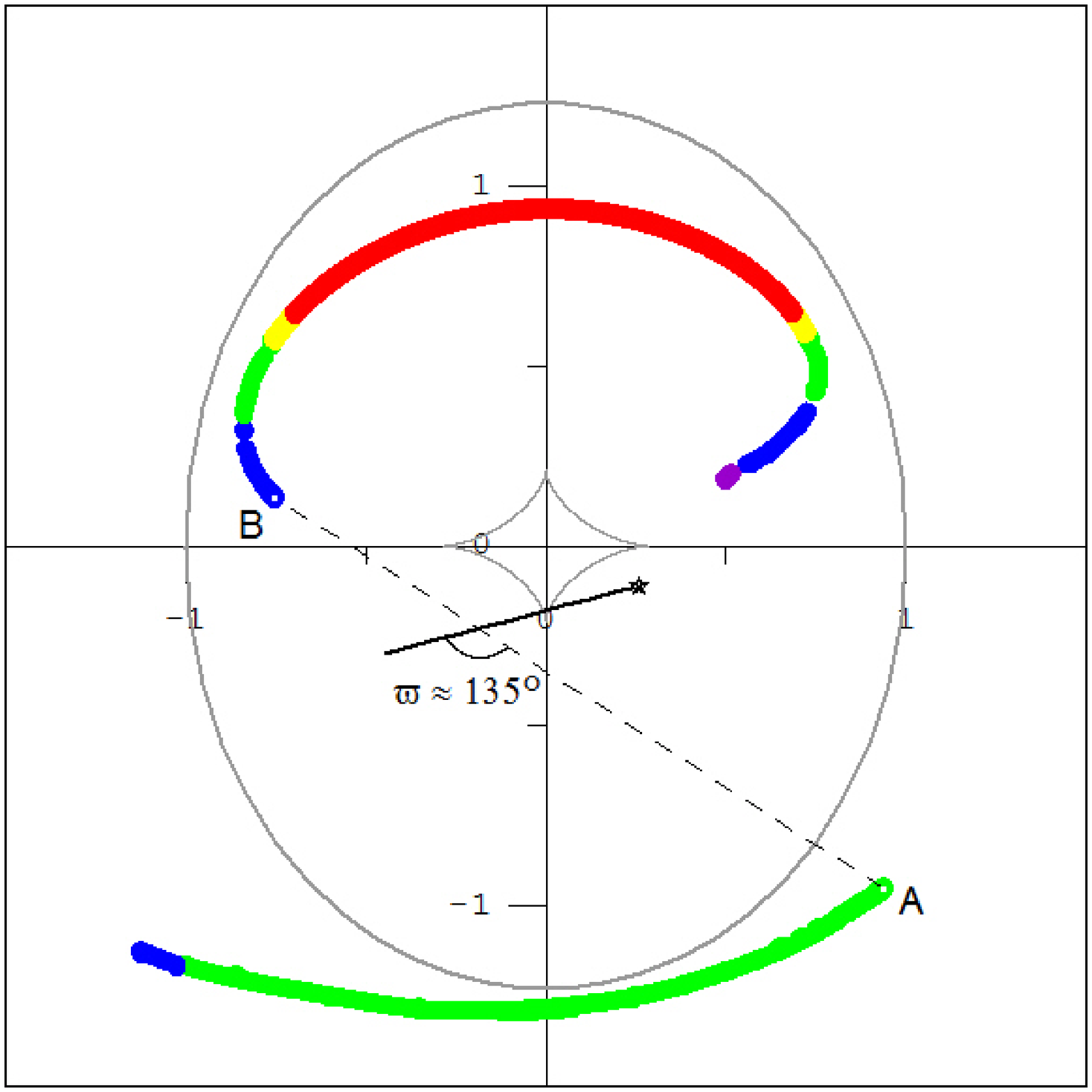}
}

\caption{({\it left}) The nearly circular image of the lensed jet in the
Model I with parameters: $q_{3d}=0.05$, $a_d=1.0$, jet slope 60°,
inclination of all components $i=30°$. ({\it right}) Results of the
B0218+357 jet modelling with the Model I.  The jet is directed at the angle
of $\varpi\simeq 135^o$ to the line connected images A and B, and cross the
tangential caustic near its bottom cusp.} \label{fig2}

\end{figure}

The system B0218+357 is a very interesting and important for astrophysical
and cosmological investigations due to several reasons: the presence of a
large scale jet and the circular-like structure in radio images; the time
delay between images of the compact core, which was measured with the high
accuracy; the significant distance from other extragalactic objects. For
this system there were obtained several set of parameters of model I and II
reproducing results of observations. An example of the mutual arrangement of
the lens, source, jet and circular-like structure is presented in
Fig.\ref{fig2} (right). In a combination with the measured time delay the
model parameters can give us estimations of the Hubble constant. It was
revealed that the considered models lead to a wide variety of its value
$H\simeq 35-90$ km s$^{-1}$ Mpc$^{-1}$ (for details see Larchenkova et
al. 2011).\\

This work was supported by the program of Russian Academy of Sciences
''Origin, Structure and Evolution of Objects in the Universe'' and
government contract P1336. AL thanks to the grant of NSh-2010.5069.2 
for the support.

\end{document}